# QED RADIATIVE CORRECTIONS FOR PARTON DISTRIBUTIONS

H. Spiesberger

*Theoretische Physik, Universität Bielefeld*
*Universitätsstraße, D–33501 Bielefeld, Germany*

## Abstract

I discuss radiative corrections due to the emission of photons from quarks which contribute to deep inelastic lepton–nucleon scattering as well as to $pp$ collisions at high energies. These corrections are dominated by quark-mass singularities which have to be absorbed into the parton distribution functions. Observable effects appear as a modification of the $Q^2$ dependence of the distribution functions. Numerical results indicate, however, that these QED corrections are negligible except at extremely large $Q^2$ and large $x$. Therefore it is safe to neglect the single and multiple photon effects in $pp$ scattering at LHC energies.

# 1 Introduction

High-energy scattering experiments with charged particles require the inclusion of electromagnetic radiative corrections due to the virtual and real emission of photons. These QED corrections are known to a high precision as far as they are related to radiation from leptons, as e.g. in the case of initial state radiation in $e^+e^-$ annihilation (see, e.g. [1]) or in the case of leptonic corrections in deep inelastic lepton–nucleon scattering [2]. In processes involving charged hadrons in the initial or final state there are QED corrections related to photon emission from hadrons or their constituents. Photons originating from hadron decays or emerging during hadronization are usually modeled in Monte Carlo programs simulating the hadronic final state and are not considered in this paper. In the following I want to discuss the effect of photon radiation from quarks entering or leaving the underlying hard scattering process. I will show that the corresponding corrections are negligible, even at very high energies.

In $e^+e^-$ annihilation into hadrons, QED corrections of the type considered here appear as photon radiation from the final state quarks. Final state radiation does not pose a severe problem since as a consequence of the KLN theorem the corresponding corrections are small. The situation is different in deep inelastic scattering or in hadron collisions. In this case there are QED corrections related to the emission of photons from quarks in the initial state. These corrections contain mass singularities related to the initial state partons. Using dimensional regularization—as usual in QCD calculations—these mass singularities appear as poles in $1/(D-4)$. If one regulates the mass singularities with the help of finite quark masses—as wide-spread in QED calculations—one finds terms proportional to $\ln m_q^2$.

In an early calculation of radiative corrections to charged current neutrino scattering by Kiskis [3], the author was worried by this fact, since it makes a large difference whether one uses $m_q = 350$ MeV (a typical constituent quark mass) leading to $\ln \mu^2/m_q^2 \simeq 10.9$ for $\mu^2 = M_W^2$, or $m_q = 5$ MeV (a typical value for the current quark mass) leading to $\ln M_W^2/m_q^2 \simeq 19.4$. Later, De Rújula, Petronzio and Savoy-Navarro [4] have argued that the unphysical dependence on quark masses can be absorbed by a redefinition of parton distribution functions. This redefinition, or *renormalization* of parton distribution functions, is well-known in the calculation of QCD radiative corrections where in complete analogy to photon radiation the emission of gluons leads to mass singularities as well.

The essential property of mass singularities is that they *factorize* [5], i.e. they can be written as a convolution of the parton-level Born cross-section $\sigma_{Born}^{parton}$ with a kernel $K(\xi, \mu^2)$ describing the effect of radiation and containing the mass-singular terms. Explicitly, for first-order corrections and omitting summation over parton types, the mass-singular part has the form

$$\frac{\alpha}{2\pi} \int_x^1 \frac{d\xi}{\xi} f(\xi) K(\xi, \mu^2) \sigma_{Born}^{parton}(\xi p). \tag{1}$$

Here $f(\xi)$ is a parton distribution function, $p$ the momentum of the incoming hadron and $\xi$ a dimensionless variable characterizing the amount of energy available for the hard scattering process after radiation. The minimum of $\xi$, $\xi_{min} = x$, is determined by the kinematics of the process. The energy scale $\mu$ (factorization scale) remains arbitrary unless the non-singular contributions are specified. Choosing $\mu$ to be a typical mass scale of the process, it is usually possible to avoid the appearance of large non-singular corrections. The lowest-order cross section is itself a convolution of parton cross sections with parton distribution functions,

$$\int_x^1 \frac{d\xi}{\xi} \delta(\xi - x) f(\xi) \sigma_{Born}^{parton}(\xi p). \tag{2}$$

Adding these contributions, one can absorb the mass-singular terms by a redefinition of the parton distribution functions:

$$\int_x^1 \frac{d\xi}{\xi} \delta(\xi - x) \int_\xi^1 \frac{du}{u} \left[ \delta(1-u) + \frac{\alpha}{2\pi} K(\xi/u, \mu^2) \right] f(\xi/u) \sigma_{Born}^{parton}(\xi p)$$
$$= \int_x^1 \frac{d\xi}{\xi} \delta(\xi - x) f^{ren}(\xi, \mu^2) \sigma_{Born}^{parton}(\xi p) \tag{3}$$

with renormalized distribution functions

$$f^{ren}(\xi, \mu^2) = \int_\xi^1 \frac{du}{u} \left[ \delta(1-u) + \frac{\alpha}{2\pi} K(\xi/u, \mu^2) \right] f(\xi/u). \tag{4}$$

The bare parton distribution functions $f(\xi)$ are, in fact, not measurable. It is rather the renormalized distributions $f^{ren}$ that have to be identified with measured (i.e. finite) quantities. The appearance of mass singularities in unphysical quantities is an artifact of the perturbative treatment. By the redefinition Eq. (4), the mass singularities disappear from the observable cross section and the renormalized distribution functions become dependent on the factorization scale $\mu^2$. The $\mu^2$ dependence is controlled by the well-known Gribov–Lipatov–Altarelli–Parisi (GLAP) equations [6]. In deep inelastic scattering they are equivalent to the renormalization group equations for the Wilson coefficients and express the fact that observable effects are independent of the energy scale at which the distribution functions are renormalized. The solution of the GLAP equations corresponds to the resummation of all powers $n$ of the leading logarithms $\alpha_s^n (\ln \mu^2)^n$.

Since mass singularities are universal, i.e. independent of the process under consideration, the definition of renormalized parton distributions is also universal. Therefore it is possible to discuss the bulk of radiative corrections in terms of parton distribution functions. This will be true if there is only one large scale in the process (e.g. in inclusive deep inelastic scattering $ep \to eX$ with $x$ not small). Then radiative corrections which are not mass-singular cannot contain large logarithms.

The above prescription for the treatment of mass singularities applies to both QCD and QED corrections. Taking into account QED corrections, the renormalization of parton distributions has to include terms due to the emission of photons from quarks, in addition to those due to the emission of gluons. Then, also the GLAP equations are modified by an additional term of the order of the electromagnetic fine-structure constant, $\alpha_e$. Apart from small non-singular contributions, the resulting modified scale dependence of parton distribution functions is the only observable effect of QED corrections in high-energy scattering of hadrons.

The modification of the GLAP evolution equations by QED corrections has also been discussed in [7]. In this work, I will present and discuss numerical results for their solutions including terms of order $\mathcal{O}(\alpha_e(\alpha_s \ln \mu^2)^n)$ for arbitrary $n$ relevant for present and future experiments like at HERA or at LHC. Moreover, I will give a simple prescription to approximately take into account the QED corrections of valence parton distributions.

The prescription to be described below applies to completely inclusive measurements of any process $h_1 h_2 \to X$, i.e. where emitted photons are not restricted by energy or angle cuts to a specific phase-space region. Experimentally this means that no attempt is made to observe emitted photons. It is a more complicated task to derive cross sections for the production of isolated hard photons (see for example [8] and references therein). In this case, mass-singular contributions can be avoided by imposing isolation cuts. For a measurement of direct photon production without isolation cuts one has to absorb part of the mass-singular contributions into photon fragmentation functions describing the non-perturbative emission of collinear photons from partons.

## 2 Formalism

After having absorbed mass-singular terms into the parton distribution functions, $q_f(x, \mu^2)$ for quarks with flavor $f$ and $G(x, \mu^2)$ for gluons, the resulting dependence on the energy scale $\mu^2$ at which the process probes the parton content of the hadrons is described by the GLAP evolution equations [6]. Using the scale variable

$$t = \ln \mu^2 / \Lambda^2, \tag{5}$$

they read

$$\frac{d}{dt} q_f(x, t) = \frac{\alpha_s(t)}{2\pi} \int_x^1 \frac{dz}{z} \left[ P_{q/q}(z, t) q_f(x/z, t) + P_{q/G}(z, t) G(x/z, t) \right], \tag{6}$$

$$\frac{d}{dt} G(x, t) = \frac{\alpha_s(t)}{2\pi} \int_x^1 \frac{dz}{z} \left[ \sum_f P_{G/q}(z, t) q_f(x/z, t) + P_{G/G}(z, t) G(x/z, t) \right]. \tag{7}$$

To leading logarithmic accuracy (LLA), the splitting kernels $P_{i/j}$ are scale independent and given by (see e.g. [10])

$$P_{q/q}(z) = C_F \left[ \frac{1+z^2}{(1-z)_+} + \frac{3}{2}\delta(1-z) \right];$$

$$P_{G/q}(z) = C_F \frac{1+(1-z)^2}{z};$$

$$P_{G/G}(z) = 2N_C \left[ \frac{1}{(1-z)_+} + \frac{1}{z} + z(1-z) - 2 \right] + \left( \frac{11}{6}N_C - \frac{1}{3}N_f \right) \delta(1-z);$$

$$P_{q/G}(z) = \frac{1}{2} \left[ z^2 + (1-z)^2 \right].$$

(8)

In the following, I restrict myself to the LLA and omit the argument $t$ in $P_{i/j}$. The running strong fine-structure constant is given by

$$\alpha_s(t) = \frac{1}{b_0 t}, \quad b_0 = \frac{33 - 2N_f}{12\pi}. \tag{9}$$

Note that the "+" distribution appearing in Eq. (8) is defined for the interval from 0 to 1:

$$\int_0^1 dz\, D_+(z) f(z) = \int_0^1 dz\, D(z) \left( f(z) - f(1) \right); \tag{10}$$

thus, if the integral is restricted to the range $x \leq z \leq 1$, an additional term $f(1)\ln(1-x)$ has to be taken into account.

The inclusion of QED corrections modifies the evolution equation for the charged parton distributions by an additional term:

$$\frac{d}{dt} q_f(x,t) = \frac{\alpha_s(t)}{2\pi} \int_x^1 \frac{dz}{z} \left[ P_{q/q}(z,t) q_f(x/z,t) + P_{q/G}(z,t) G(x/z,t) \right]$$
$$+ \frac{\alpha_e(t)}{2\pi} \int_x^1 \frac{dz}{z} P^\gamma_{q/q}(z,t) q_f(x/z,t) \tag{11}$$

where

$$P^\gamma_{q/q}(z) = e_f^2 \left[ \frac{1+z^2}{(1-z)_+} + \frac{3}{2}\delta(1-z) \right] = \frac{e_f^2}{C_F} P_{q/q}. \tag{12}$$

Here the running electromagnetic fine-structure constant appears, which is given by

$$\alpha_e(t) = \frac{\alpha(0)}{1 - \frac{\alpha}{3\pi} \sum_f e_f^2 (t - t_{m_f}) \theta(t - t_{m_f})}, \tag{13}$$

with $t_{m_f} = \ln(m_f^2/\Lambda^2)$ and $\alpha(0) = 1/137.036\ldots$. $e_f$ are the fermion charges in units of the positron charge and the fermion-mass thresholds have been approximated by the step function $\theta(t - t_{m_f})$.

It convenient to use the following combinations of parton distribution functions:

$$U(x) = \sum_{gen} (u(x) + \bar{u}(x)),$$
$$D(x) = \sum_{gen} \left( d(x) + \bar{d}(x) \right),$$
$$\Sigma(x) = U(x) + D(x),$$
$$\Delta(x) = U(x) - D(x),$$

(14)

where $\sum_{gen}$ means summation over the generations. Then, denoting the convolution of splitting kernels with distribution functions by the symbol $\otimes$ and the derivative with respect to the scale variable $t$ by a dot, the evolution equations can be written as:

$$\dot{\Sigma} = \frac{\alpha_s}{2\pi} P_{q/q} \otimes \Sigma + 2N_f \frac{\alpha_s}{2\pi} P_{q/G} \otimes G + \frac{\alpha_e}{2\pi} P^\gamma_{q/q} \otimes \left(\frac{5}{18}\Sigma + \frac{1}{6}\Delta\right),$$

$$\dot{\Delta} = \frac{\alpha_s}{2\pi} P_{q/q} \otimes \Delta + \frac{\alpha_e}{2\pi} P^\gamma_{q/q} \otimes \left(\frac{1}{6}\Sigma + \frac{5}{18}\Delta\right), \tag{15}$$

$$\dot{G} = \frac{\alpha_s}{2\pi} P_{G/q} \otimes \Sigma + \frac{\alpha_s}{2\pi} P_{G/G} \otimes G.$$

In contrast to the pure QCD evolution, the non-singlet combination of parton distributions, $\Delta(x)$, does not decouple from the singlet sector due to the different charges of up- and down-type quarks. For our purpose it is convenient to separate the solution of the equations without the QED term, i.e. to write

$$\Sigma = \Sigma^0 + \sigma, \quad \Delta = \Delta^0 + \delta, \quad G = G^0 + g \tag{16}$$

where $\Sigma^0$, $\Delta^0$, and $G^0$ obey the evolution equations Eq. (15) with $\alpha_e = 0$. $\sigma$, $\delta$, and $g$ are corrections of relative order $\mathcal{O}(\alpha_e)$. Neglecting terms of order $\mathcal{O}(\alpha_e^2)$, one obtains equations for the QED contributions to the parton distributions:

$$\dot{\sigma} = \frac{\alpha_s}{2\pi} P_{q/q} \otimes \sigma + 2N_f \frac{\alpha_s}{2\pi} P_{q/G} \otimes g + \frac{\alpha_e}{2\pi} P^\gamma_{q/q} \otimes \left(\frac{5}{18}\Sigma + \frac{1}{6}\Delta\right),$$

$$\dot{\delta} = \frac{\alpha_s}{2\pi} P_{q/q} \otimes \delta + \frac{\alpha_e}{2\pi} P^\gamma_{q/q} \otimes \left(\frac{1}{6}\Sigma + \frac{5}{18}\Delta\right), \tag{17}$$

$$\dot{g} = \frac{\alpha_s}{2\pi} P_{G/q} \otimes \sigma + \frac{\alpha_s}{2\pi} P_{G/G} \otimes g.$$

Note that although the gluons do not couple directly to photons, their distribution function is modified through higher-order contributions induced by their coupling to the quarks. The correction $g$ is formally of order $\mathcal{O}(\alpha_e \alpha_s)$. In Eqs. (17) the full $\alpha_s$ evolution of the QED contributions is kept.

## 3 Numerical Results

The equations Eqs. (17) lend themselves directly to an iterative numerical solution, given the QCD evolved solutions $\Sigma^0(x,t)$ and $\Delta^0(x,t)$. The use of Eqs. (15) would require initial conditions for $q_f(x,t_0)$ and $G(x,t_0)$ at some reference scale $t_0 = \ln(Q_0^2/\Lambda^2)$. High precision and a stable numerical algorithm is needed if the small corrections of $\mathcal{O}(\alpha_e)$ were to be determined directly from Eqs. (15) instead of Eqs. (17). Therefore it is preferable to use Eqs. (17). I checked my algorithm by comparing the solutions from these two equations. As an additional check, the algorithm was used to solve Eqs. (15) with $\alpha_e = 0$ and initial conditions $q_f(x,t_0)$ taken from one of the commonly used parametrizations of parton distribution functions, as provided for example by the program library PAKPDF [11]. The resulting $Q^2$ dependence of the pure QCD evolution was then compared with the $Q^2$ dependence of the corresponding parametrization. In view of the fact that the available parametrizations are only approximate solutions of the GLAP equations having their own limited precision, the agreement between these different methods was remarkably good.

The solution of the differential equations Eqs. (15) or (17) requires initial conditions, i.e. the knowledge of parton distribution functions at a reference scale $\mu^2 = Q_0^2$. The input distribution functions $q_f(x, Q_0^2)$, $G(x, Q_0^2)$ have to be taken from experimental data which (with only a few exceptions) have not been corrected for radiative corrections due to photon emission from quarks. Consequently, the QED corrections $\sigma$, $\delta$, and $g$ in Eqs. (17) have to vanish at the reference scale.

In Figs. 1, 2, and 3, I show results for the corrections to the distribution functions $U(x, Q^2)$, $D(x, Q^2)$, $G(x, Q^2)$ and the structure function

$$F_2^p(x, Q^2) = x \left[ \frac{5}{18} \Sigma(x, Q^2) + \frac{1}{6} \Delta(x, Q^2) \right] \tag{18}$$

measured in deep inelastic electron–proton scattering at HERA. The figures show the QED corrections in per cent relative to the distribution functions obtained from the GLAP equations without the QED term. The input distributions were taken from [12] (set 1.1) since those parametrizations are fairly simple and the numerical solution of Eqs. (17) requires less computer time than for other parton distributions. The figures show small, negative corrections at the per-mille level for all values of $x$ and $Q^2$ relevant in forthcoming experiments. Only at large $x \gtrsim 0.5$ and large $Q^2 \gtrsim 10^3$ GeV$^2$ the corrections reach the magnitude of one per cent.

The increase of corrections for $x \to 1$ is due to the $\ln(1-x)$ terms appearing in the evaluation of the "+" distributions. Although large relative corrections are reached in this limit, they are of no practical relevance, since parton distributions and cross sections decrease as $x \to 1$ and to reach an experimental accuracy of a few per cent at large $x$ will be very unlikely.

The largest corrections are obtained for up-type quarks due to the larger charge factor 4/9 as compared to 1/9 for down-type quarks. Therefore the correction to $F_2^p$ is close to that for $U(x)$. The gluon distribution, being of order $\mathcal{O}(\alpha_e \alpha_s)$, is corrected by less than 0.1% up to values of $x$ of about 0.2.

Vice versa, the effect of $g$ being non-zero on the quark distributions is of the same size. Fixing $g$ at $g(x) \equiv 0$ would lead to different results, particularly at small $x$, though at the same level of a few per mille and with the same cross features. At the level of the observed QED corrections $\sigma, \delta$, and $g$, their QCD evolution is non-negligible.

The corrections vanish for $Q^2 \to Q_0^2$ since I assumed that the input distributions $q_f(x, t_0)$ and $G(x, t_0)$ have been extracted from experiment at the reference scale $Q_0^2$ without subtracting quarkonic QED corrections.

The asymptotic behavior for $x \to 0$ can be checked analytically. It is well-known that a singular behavior of distributions $\propto x^{-\alpha}$ for $x \to 0$ remains unchanged by the GLAP equations if $\alpha > 1$. Thus the $\mathcal{O}(\alpha_e)$ corrected distributions have the same power behavior as the uncorrected ones, the ratio consequently reaching a constant value for $x \to 0$. The value of this constant depends on the power $\alpha$ and other details of the distribution functions. The valence parts of $U(x)$ and $D(x)$, however, which vanish at $x = 0$, receive positive corrections at small $x$, thus producing the well-known physical picture: radiation of gluons as well as of photons leads to a depletion at large $x$ and an enhancement at small $x$, i.e. partons are shifted to smaller $x$.

The corrections strongly depend on the input distribution functions; differences at the per-mille level are found. But still these differences are irrelevant when compared with the expected experimental precision of structure-function measurements.

For valence distributions there is a simple prescription to include the effect of QED radiation. Using as variable the evolution length (in leading order)

$$\xi_s(t) = \int_{t_0}^{t} dt' \frac{\alpha_s(t')}{2\pi} = \frac{1}{2\pi b_0} \ln \frac{t}{t_0}, \tag{19}$$

the GLAP equation for valence quark distributions has the simple form

$$\frac{d}{d\xi_s} q_{val}(x, \xi_s) = \int_x^1 \frac{dz}{z} P_{q/q}(z, \xi_s) q_{val}(x/z, \xi_s). \tag{20}$$

Then, the substitution

$$\begin{aligned}\xi_s(t) \to \xi_{s+e}(t) &= \int_{t_0}^{t} dt' \left( \frac{\alpha_s(t')}{2\pi} + \frac{e_f^2}{C_F} \frac{\alpha_e(t')}{2\pi} \right) \\ &= \xi_s(t) + \frac{3e_f^2}{2C_F \sum_f e_f^2} \ln \frac{\alpha_e(t)}{\alpha_e(t_0)}, \end{aligned} \tag{21}$$

automatically takes account of QED radiative effects. This means, the evolution length is increased by QED radiation. Given the solution $q_{val}(x, t)$ of Eq. (20), one obtains the solution of the GLAP equation including QED effects by

$$q_{val}(x, t) \to q_{val}(x, \bar{t}) \tag{22}$$

with

$$\bar{t} = t \left( \frac{\alpha_e(t)}{\alpha_e(t_0)} \right)^{3\pi e_f^2 b_0 / (C_F \sum_f e_f^2)}. \tag{23}$$

This simple prescription is sufficient for including quarkonic QED corrections into the structure functions at large $x$ where valence contributions dominate.

## 4 Concluding Remarks

I have shown that radiation of photons from quarks leads to negligible corrections of cross sections for high-energy scattering processes involving hadrons. The leading effect is a modification of the scale dependence of parton distribution functions described by a modification of the GLAP evolution equations. These equations include the effect of multiple photon emission as far as it is enhanced by a logarithm of the energy scale of the process. Non-leading corrections have not been calculated, but they are of the order of $\mathcal{O}(\alpha_e/\pi) \simeq 0.2\%$ and therefore also negligible.

A calculation of QED corrections using definite values for quark masses [13] and not factorizing the corresponding mass-singular contributions is misleading and usually leads to a gross overestimation of the effect.

I did not discuss the question of how to model the production of soft or collinear photons in the final state of a hard scattering process. This should be viewed as a part of the non-perturbative hadronization process. Commonly used Monte Carlo programs apply a two-step procedure where in a first step a parton cascade is developed. This cascade obeys the GLAP evolution and may allow for photon emission in addition to gluon radiation. The termination of the parton cascade requires a cut-off which could be implemented in a reasonable way by using effective quark masses. Usually those quark masses are identified with the constituent quark masses. It is important to understand, however, that this prescription does not correct the hard scattering cross section. The purpose of these Monte Carlo programs is to provide a model for the distribution of energy and momentum among final state hadrons.

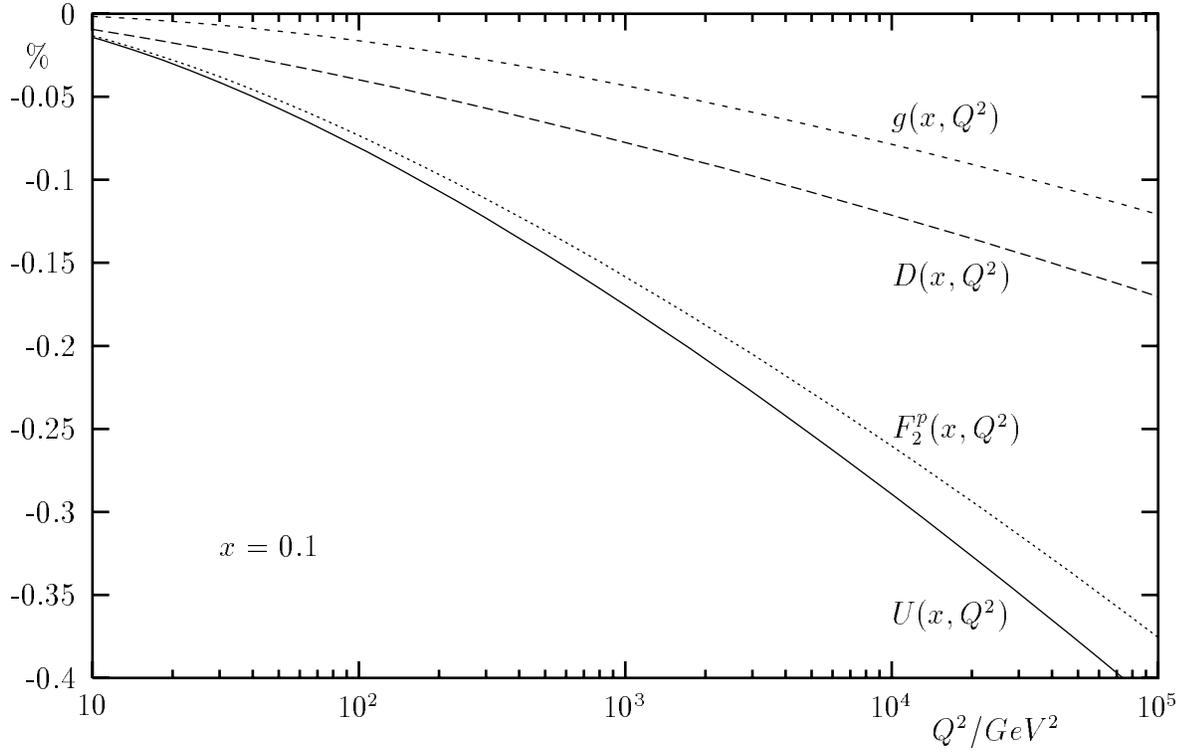

Figure 1: $Q^2$ dependence of the QED corrections (in per cent, see text) to parton distributions and the structure function $F_2^p$ for deep inelastic lepton–proton scattering at $x = 0.1$. Input parton distributions were taken from [12].

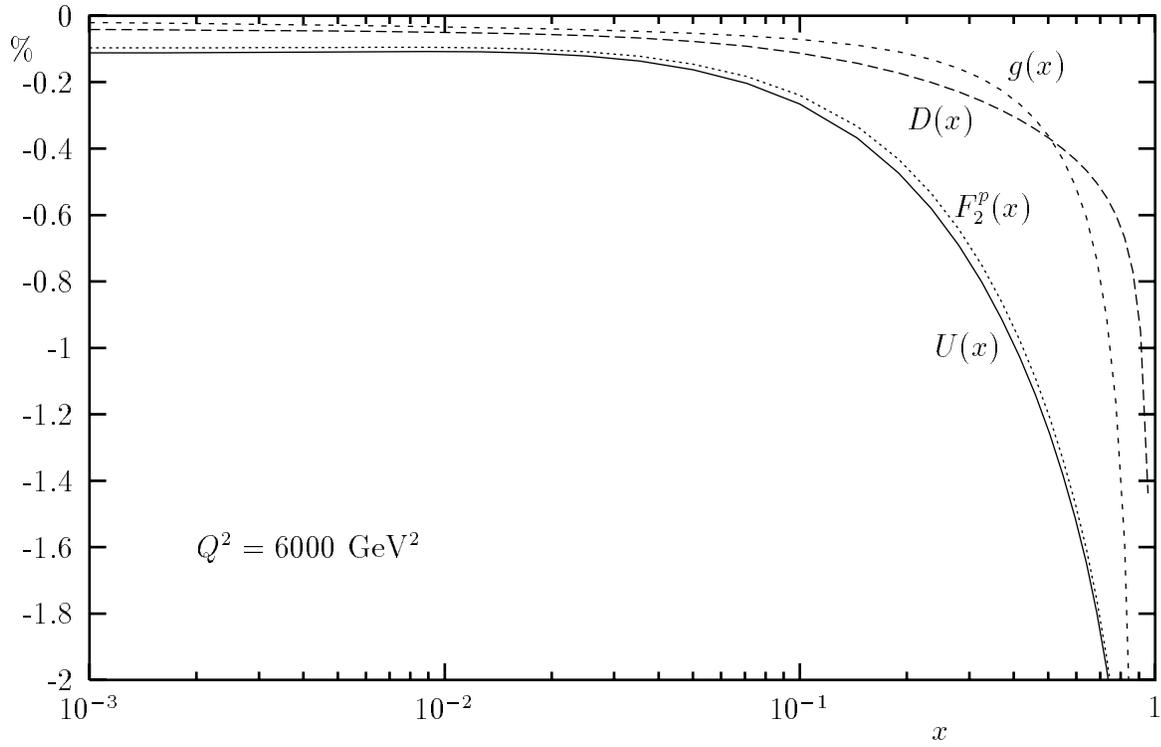

Figure 2: $x$ dependence of the QED corrections (in per cent, see text) to parton distributions and the structure function $F_2^p$ for deep inelastic lepton–proton scattering at $Q^2 = 6 \times 10^3$ $GeV^2$. Input parton distributions were taken from [12].

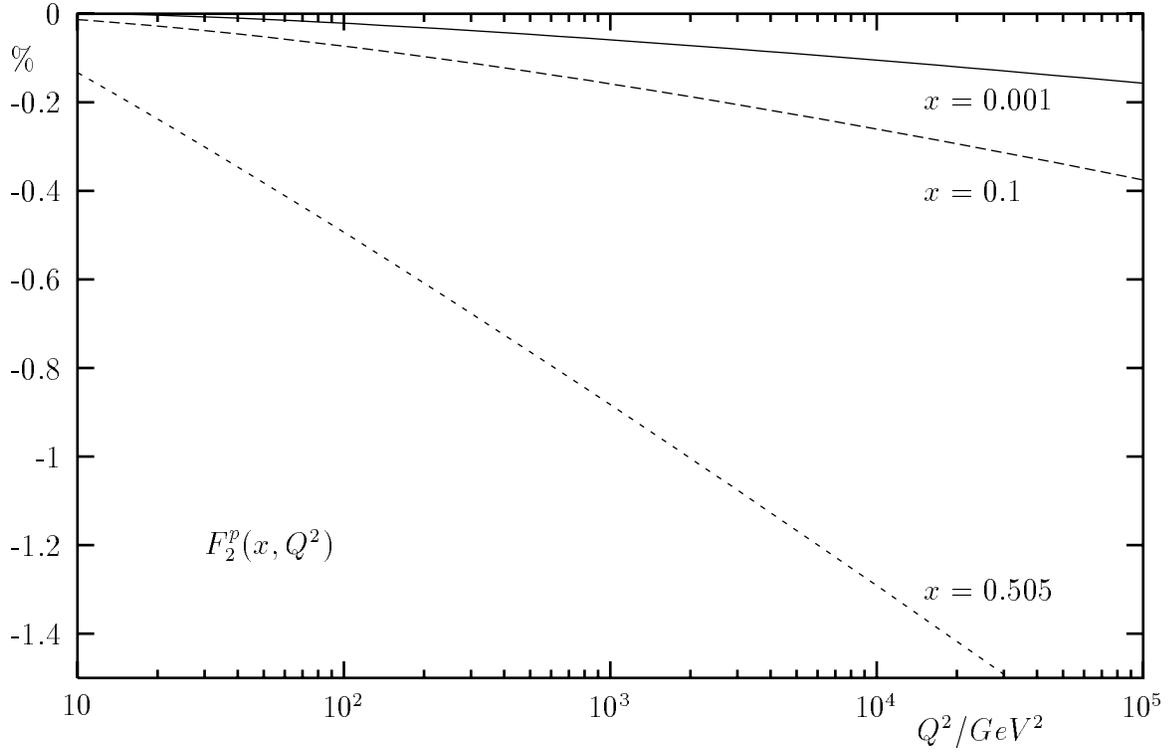

Figure 3: $Q^2$ dependence of the QED corrections (in per cent, see text) to the structure function $F_2^p$ for deep inelastic lepton–proton scattering at $x = 0.001$, $x = 0.1$ and $x = 0.505$. Input parton distributions were taken from [12].